\documentclass{emulateapj}

\shorttitle{Are red halos made of low-mass stars?}
\shortauthors{Zackrisson \& Flynn}

\begin{document}

\title{Are the red halos of galaxies made of low-mass stars?\\Constraints from subdwarf star counts in the Milky Way halo}
\author{Erik Zackrisson\altaffilmark{1,2,3}$^*$, \& Chris Flynn\altaffilmark{1}}
\altaffiltext{*}{E-mail: ez@astro.su.se}
\altaffiltext{1}{Tuorla Observatory, Department of Physics and Astronomy, University of Turku, V\"ais\"al\"antie 20, FI-21500 Piikki\"o, Finland}
\altaffiltext{2}{Stockholm Observatory, AlbaNova University Center, 106 91 Stockholm, Sweden}
\altaffiltext{3}{Department of Astronomy and Space Physics, Box 515, 751 20 Uppsala, Sweden}

\begin{abstract}

Surface photometry detections of red and exceedingly faint halos around
galaxies have resurrected the old question of whether some non-negligible fraction
of the missing baryons of the Universe could be hiding in the form of faint,
hydrogen-burning stars. The optical/near-infrared colours of these red halos
have proved very difficult to reconcile with any normal type of stellar
population, but can in principle be explained by advocating a bottom-heavy
stellar initial mass function. This implies a high stellar mass-to-light
ratio and hence a substantial baryonic mass locked up in such halos. Here, we explore
the constraints imposed by current observations of ordinary stellar halo
subdwarfs on a putative red halo of low-mass stars around the Milky
Way. Assuming structural parameters similar to those of the red halo recently
detected in stacked images of external disk galaxies, we find that a smooth
halo component with a bottom-heavy initial mass function is completely ruled out by current star
count data for the Milky Way. All viable smooth red halo models with a density
slope even remotely similar to that of the stacked halo moreover contain far
too little mass to have any bearing on the missing-baryon problem. However, we
note that these constraints can be sidestepped if the red halo stars are locked
up in star clusters, and discuss potential observations of other nearby
galaxies that may be able to put such scenarios to the test.

\end{abstract}





\section{Introduction} 
The quest to unravel the nature of dark matter, estimated to make up around
90\% of the total matter content \citep[e.g.][]{Komatsu et al.}, remains one of
the most important tasks of modern cosmology. Dark matter appears to exist
in at least two separate forms: one baryonic, and one non-baryonic. While the
non-baryonic component is the dominant one, a substantial fraction of the
baryons in the low-redshift Universe \citep[$\approx
1/3$--2/3;][]{Fukugita,Fukugita & Peebles a,Nicastro et al.,Prochaska & Tumlinson} are also at
large. 

These missing baryons could in principle be hiding in a variety of
different forms: as faint/failed stars or stellar remnants \citep[so-called
Massive Astrophysical Compact Halo Objects or MACHOs;][]{Griest}, as cold gas
clouds \citep{Pfenniger et al.,Pfenniger & Combes}, as a warm/hot intergalactic
medium \citep{Cen & Ostriker,Davé et al.} or as hot gaseous halos around
galaxies \citep{Maller & Bullock,Fukugita & Peebles b,Sommer-Larsen}. While
current simulations seem to favour the latter two alternatives as the main
reservoirs, observations are still unable to confirm this hypothesis
\citep[see][for a review]{Bregman}.

The old idea of baryonic dark matter in the form of faint, low-mass stars has
recently gained new momentum through surface photometry detections of very red
and exceedingly faint structures -- ``red halos'' -- around galaxies of
different types. The history of this topic goes back to the mid-90s, when deep
optical and near-IR images indicated the presence of a faint halo around the
edge-on disk galaxy NGC 5907 \citep[e.g.][]{Sackett et al.,Lequeux et al. a,Rudy
et al.,James & Casali}. The colours of this structure were much too red
to be reconciled with any normal type of stellar population, and indicative of
a halo population with an abnormally high fraction of low-mass stars. At around
the same time, \citet{Molinari et al.} also announced the detection of a red
halo around the cD galaxy at the centre of the galaxy cluster Abell
3284. Skepticism grew with the discovery of what appeared to be the remnants of
a disrupted dwarf galaxy close to NGC 5907 \citep{Shang et al.}, leading to
suggestions that this feature, in combination with other effects, could have
resulted in a spurious halo detection \citep{Zheng et al.}. While follow-up
observations with the Hubble Space Telescope (HST) took some of the edge out of
this criticism \citep{Zepf et al.}, the field fell into disrepute. Deeper images have later revealed a wealth of tidal streams in the halo of NGC 5907 \citep{Martinez-Delgado et al.}, but this does not by itself explain the red excess originally detected.

The red halos would not die quietly though, and new reports started to surface
a few years later. First, \citet{Bergvall & Östlin} and \citet{Bergvall et al.}
presented deep optical/near-IR images of faint and abnormally red structures
around blue compact galaxies. \citet*{Zibetti et al.} then stacked images
of 1047 edge-on disk galaxies from the Sloan Digital Sky Survey (SDSS) and
detected a halo population with a strong red excess and optical colours
curiously similar to those previously derived for NGC 5907 -- again very
difficult to reconcile with standard halo populations. The halo detected
around an edge-on disk galaxy at redshift $z=0.322$ in the Hubble Ultra Deep
Field shows similarly red colours \citep{Zibetti & Ferguson}, and \citet{Tamm
et al.} argue that even the halo of Andromeda displays a pronounced red excess.

\citet{Zackrisson et al. a} analyzed the colours of some of these new
detections and found that the halos of both blue compact galaxies and stacked edge-on
disks could be explained by a stellar population with a very bottom-heavy
initial mass function ($dN/dM\propto M^{-\alpha}$ with $\alpha\approx
4.50$). The high mass-to-light ratio of such a population makes it a potential
reservoir for at least part of the baryons missing from current
inventories. While the stellar initial mass function (IMF) is often assumed to
be universal, recent observational studies suggest that it may vary both as a
function of environment \citep{Hoversten & Glazebrook}, and as a function of
cosmic time \citep{van Dokkum}. Corroborating evidence for an IMF as extreme as
that advocated by \citet{Zackrisson et al. a} also comes from star counts in
the field population of the LMC, where a slope of $\alpha\approx 5$--6 was
derived for masses $\geq 1 \ M_\odot$ \citep{Massey,Gouliermis et al.}.

Taking the red halo detections at face value, one may ask whether the Milky Way
itself could be surrounded by a hitherto undetected red halo of low-mass,
hydrogen-burning stars, with photometric properties similar to the halo
detected around stacked edge-on disks. The known baryonic components
(thin disc, thick disk, bulge and standard stellar halo) of the Milky Way
contribute around 5--$6\times 10^{10}\ M_\odot$ \citep[e.g.][]{Sommer-Larsen &
Dolgov,Klypin et al.,Flynn et al.} to the Milky Way's virial mass of $\approx
1\times 10^{12} \ M_\odot$ \citep[][]{Klypin et al.}. A cosmic baryon fraction
of $\Omega_\mathrm{baryons}/\Omega_\mathrm{M}\approx 0.17$ \citep{Komatsu et
al.}, combined with the theoretical prediction that the baryon fraction should
be $\approx 90\%$ of the cosmic average for a Milky Way-sized halo \citep{Crain
et al.}, on the other hand suggests the presence of some $\approx 1.5\times
10^{11}\ M_\odot$ of baryonic material within its virial radius, leaving at
least $60\%$ of its baryons to be found.

Low-mass stars would in principle be detectable through microlensing effects,
and such putative MACHOs may already have been discovered in the halos of both
the Milky Way and M31 \citep[e.g.][]{Alcock et al.,Calchi Novati et
al. a,Riffeser et al.}. Their masses (0.1--1 $M_\odot$) and inferred contribution to the mass
of the dark matter of galaxies ($\approx 20 \%$) are, however, very difficult
to reconcile with any kind of stellar MACHO candidate
\citep[e.g.][]{Freese}. This result, coupled to the fact that competing teams
have failed to confirm these detections \citep[][]{Tisserand et al.,de Jong et
al.} have led to the suspicion that the observations must have been
misinterpreted \citep[e.g.][]{Belokurov et al.} or that the compact objects
detected through this technique may be of non-baryonic origin (e.g. primordial
black holes, mirror matter objects, preon stars or scalar dark matter
miniclusters -- see \citealt{Zackrisson & Riehm} for a more thorough
discussion).

If some of the missing baryons of the Milky Way are locked up in the form of
hydrogen-burning stars in a red halo, such a structure must also have evaded
the faint star counts aimed to constrain the luminosity function of halo
subdwarfs \citep[e.g.][]{Gould et al., Gould,Digby et al.,Brandner}, since no
significant excess of low-mass stars has yet been detected with this method. In
fact, direct observations of this kind are expected to impose much stronger constraints on
main sequence stars in the halo than what current
microlensing surveys can achieve.

In this paper, we explore to what extent a smooth red halo of low-mass stars
similar to that detected by \cite{Zibetti et al.} might already be ruled out by
these observations of halo subdwarfs in the Milky Way. In section 2, we
describe the observational data used. Section 3 outlines the method used to
test red halo models against these observations. The resulting constraints on
red halo models are presented in section 4. Section 5 discusses the robustness
of these constraints and section 6 summarizes our findings.

\section{Observational data}
\subsection{Surface photometry}
Using stacked SDSS images of 1047 edge-on galaxies, \citet[][hereafter
Z04]{Zibetti et al.} derived the colours of a diffuse halo in a
wedge-shaped region located at a projected distance of $R_\mathrm{proj}=2\times
r_\mathrm{exp}$ from the centre of the stacked disk, where $r_\mathrm{exp}$
corresponds to the exponential scale length of the disk. While the $g-r$ colour
of this region ($g-r=0.65\pm 0.1$) is similar to those of old stellar
populations like globular clusters or elliptical galaxies, $r-i$ is
anomalously red ($r-i=0.60\pm0.1$) and difficult to reconcile with any known
type of stellar population. These colours can, however, be explained in the
framework of a stellar population with an IMF slope of $\alpha=4.50$ \citep{Zackrisson et al. a}, i.e. a stellar halo overly abundant in low-mass stars. One the other hand, the wavelength-dependence of the far wings of the point-spread function (PSF) may introduce spurious colours in the outskirts of extended objects \citep[e.g.][]{Michard,Sirianni et al.}. Recently, \citet{de Jong} has argued that Z04 may have underestimated the effects of the SDSS PSF, and that the reported $r-i$ halo colour therefore suffers from artificial reddening due to scattered light. At the current time, it is very difficult to assess whether this accounts for all of the red excess, or just some part thereof. With this in mind, the IMF slope of $\alpha=4.50$ derived by \citet{Zackrisson et al. a} is likely to represent an upper limit, barring systematic uncertainties in the fit due to lingering problems with current models for low-mass stars \citep[e.g.][]{Casagrande et al.}. In what follows, we will use the $i$-band data of Z04, but allow for the possibility that both the overall surface brightness level of the halo and the colour may have been overestimated. 

When comparing this average halo to the Milky Way, we have adopted a scale
length for the Milky Way's disk of $r_\mathrm{exp}=2.5$ kpc, in agreement with
recent estimates based on both optical and near-infrared data \citep{Gardner et al.}. The measured $i$-band surface brightness level
of the wedge in the stacked frame is $\mu_{\mathrm{red},\; i}\approx 26.7$ mag
arcsec$^{-2}$, but we also explore the consequences of putative red halos with
surface brightness levels both brighter and fainter than this. There are
several reasons for this strategy. Firstly, part of the light measured at this
distance is likely to come from the disk and the far wings of the SDSS
PSF, implying that the red halo should be somewhat {\it
fainter} at this distance than suggested by the surface brightness level
actually measured. Secondly, the analysis presented by Z04 indicates that the
surface brightness of the halo scales with the luminosity of the disk, which
would suggest a red halo {\it brighter} than $\mu_{\mathrm{red},\; i}\approx
26.7$ at $R_\mathrm{proj}=2\times r_\mathrm{exp}$ for a relatively luminous galaxy
like the Milky Way. On the other hand, it is not clear what the distribution of
red halo properties within the stacked sample is, and individual differences
between galaxies may possibly compensate for such a trend. Here, we therefore
consider the possibility that a hypothetical red halo surrounding the Milky Way
could be substantially different than the cosmic average as derived by Z04.

While Z04 derive a halo flattening of $q\approx 0.6$ and a density profile with
power-law slope $\beta\approx 3$ for the $i$-band, we here explore the
consequences of red halos with $q=0.5$--1.5 and power $\beta=0$--10. This very
generous range of parameter values ensure that the constraints derived are
conservative, in the sense that they allow any hypothetical red halo maximal
leverage.
\begin{figure}[t]
\centering
\plotone{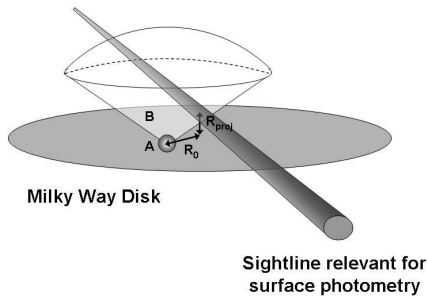}
\caption{Schematic illustration of the relevant data situation. The volumes
probed by the subdwarf observations of \citet{Digby et al.} and \citet{Gould et
al.} are indicated by the gray sphere marked A and the cone marked B, respectively. $R_0$ represents the distance from the Galactic centre to the
position of the Sun. The sightline along which the red halo colours have been
measured through surface photometry around stacked SDSS disks is indicated by
the gray cylinder, located at a projected distance of $R_\mathrm{proj}$ above
the plane of the disk.
\label{schematic}}
\end{figure}

\subsection{Star counts}
Subdwarfs are main sequence stars in the mass range $\approx 0.08$--$1 \
M_\odot$ that because of their low metallicities have lower $V$-band
luminosities than disk stars of the same colour. These objects make up the bulk
of stars in the hitherto detected stellar halo of the Milky Way, and have been
used in numerous studies to constrain the characteristics of this
structure. Since adopting a bottom-heavy IMF of the \citet{Zackrisson et al. a}
type would boost the fraction of subdwarfs in a population, it makes sense to
use the observed number densities of these objects to constrain such scenarios.

Large samples of halo subdwarfs can be collected using two different
techniques: proper-motion selection and colour-magnitude selection in deep
fields.

The first method exploits the high heliocentric velocities statistically
expected from stars belonging to the halo and selects candidate halo stars by
imposing a minimum proper motion limit on the sample \citep[e.g.][]{Gould,Digby
et al.}. A kinematical model is then used to correct the resulting statistics
for incompleteness and contamination by thin and thick disk stars. This method
is currently limited to halo stars within a few kpc from the position of the
Sun.

The second method is based on identification of stars in long exposures (``deep
fields'') of the sky at high Galactic latitudes \citep{Gould et al.,Brandner}
and the use of colour-magnitude criteria to reject objects in the disk. While
this technique probes subdwarfs at much fainter magnitude limits (and thereby
larger distances) than the former, the selection criteria prevent it from
probing subdwarfs within a few kpc of the Sun.

The volumes probed by these two methods are complementary, with almost no
overlap. Curiously enough, the scaling of the luminosity function of halo
subdwarfs derived by these two techniques differ somewhat \citep[by a factor of
2--3; see][for a comparison]{Digby et al.}. The exact reason for this
discrepancy is not well-understood, but could possibly be due to a difference
in the characteristics of the inner and outer halo. Indeed, \citet{Carollo et
al.} recently found strong evidence for two separate structural components in
the Galactic halo, with the inner halo being substantially more flattened than
the outer.

Here we make no attempt to reconcile the measurements resulting from these
different techniques. Instead, we assume that the subdwarf luminosity
functions and the halo parameters derived by the two techniques are correct in
the mutually exclusive volumes for which they are relevant. In what follows, we
therefore consider two sets of constraints, A and B, based on the
proper-motion samples of by \citet{Digby et al.} and the deep-field samples of
\citet{Gould et al.}, respectively. Even though there are many large studies
based on the first technique, the differences between these and those of
\citet{Digby et al.} are minor and will not have any significant impact on the
current study. While there are slight differences between the assumptions made
in \citet{Gould et al.} and \citet{Digby et al.} about the functional form of
density profile of the stellar halo, both studies are able to produce good fits
to the number of subdwarfs observed in the volumes probed, and this is what
matters for the current study.

\citet{Digby et al.} assume a density profile for the stellar halo of the
form\footnote{Here, we have corrected an obvious misprint in \citet{Digby et
al.}, related to the definition of $\beta$.}:

\begin{equation}
n_\mathrm{A}(x,y,z)=n_\mathrm{A,\odot}\left( \frac{x^2+y^2+(z/q)^2 +
R_\mathrm{c}^2}{R_0^2+ R_\mathrm{c}^2} \right)^{-\beta/2},
\end{equation} 
where $x$, $y$, $z$ are galactocentric coordinates. $n_\mathrm{\odot}$ is the number density of subdwarfs in the vicinity of
the Sun, $q$ is the halo flattening parameter, $R_0$ is the distance from the
Sun to the centre of the Milky Way (throughout this paper assumed to be 8.0
kpc), $R_\mathrm{c}$ is a core radius (assumed to be $R_\mathrm{c}=1.0$ kpc)
and $\beta$ is the exponent of the density power-law. While \citet{Digby et al.} are unable to
impose useful constraints on $q$, they find a best-fitting $\beta=3.15$ and
adopt $q=0.55$ in their plots. In what follows, these values will be adopted by
us as well.

\citet{Gould et al.} instead assume that the subdwarfs of the stellar halo are
distributed according to a density profile of the form:

\begin{equation}
n_\mathrm{B}(x,y,z)=n_\mathrm{B,\odot}\left( \frac{x^2+y^2+(z/q)^2}{R_0^2} \right)^{-\beta/2}. 
\end{equation} 
where $x$, $y$, $z$ are galactocentric coordinates. With this assumption, they derive best fitting values of $\beta=3.13$ and $q=0.82$.

To derive $n_\mathrm{A,\odot}$ and $n_\mathrm{B,\odot}$, we need to adopt
a luminosity range for subdwarfs. For simplicity, we consider all main sequence
stars that fall in the luminosity range probed by {\it both} \citet{Digby et
al.} and \citet{Gould et al.} to be subdwarfs, i.e. objects with $V$-band
luminosities $7.5\lesssim M_V\lesssim 12.4$. This then implies
$n_\mathrm{A,\odot}\approx 1.7\times 10^5$ kpc$^{-3}$ and
$n_\mathrm{B,\odot}\approx 6.4\times 10^4$ kpc$^{-3}$.

\section{Star counts confront surface photometry}
Here we use the best-fitting halo parameters derived by \citet{Digby et
al.} and \citet{Gould et al.} to compute the mean number densities of subdwarfs
$\overline{n_\mathrm{A}}$ and $\overline{n_\mathrm{B}}$ within the relevant
volumes $V_\mathrm{A}$ and $V_\mathrm{B}$:
\begin{equation}
\overline{n_\mathrm{A/B}}=\frac{1}{V_\mathrm{A/B}}\int n_\mathrm{A/B}(x,y,z) \;
\mathrm{d}V_\mathrm{A/B}.
\end{equation} 

Any additional, smooth halo population that involves higher subdwarf densities
than $\overline{n_\mathrm{A}}$ or $\overline{n_\mathrm{B}}$ in these volumes
should already have been detected in either the \citet{Digby et al.} or
\citet{Gould et al.} surveys and can therefore be ruled out. By simply checking
$\overline{n_\mathrm{A}}$ and $\overline{n_\mathrm{B}}$ against the predictions
of various red halo models, we can therefore place conservative -- yet very
strong -- constraints on any hitherto undetected, smooth red halo of low-mass
stars around the Milky Way.

The volumes A and B considered relevant for the studies by
\citet{Digby et al.} and \citet{Gould et al.}  are illustrated in
Fig.~\ref{schematic}. For simplicity, we take the volume covered by
\citet{Digby et al.} to be a sphere with radius 2.8 kpc centered on the
position of the Sun. In the case of \citet{Gould et al.}, the actual volumes
probed correspond to an ensemble of many very narrow cones of different length
(given by the flux limits of the HST fields used), with their tips removed to
avoid contamination from stars belonging to the disk. Here, we approximate
these volumes by a single, wide cone with opening angle 60$^\circ$ and height
40 kpc, with the tip (within 2.3 kpc of the plane) removed. In
Fig.~\ref{schematic}, we have only depicted one cone, whereas in reality, stars
were selected from both directions away from the Milky Way disk. Since we are
only considering halo models that are symmetric with respect to the disk, this
has no impact on our results.  Using $V_\mathrm{A}\approx 92$ kpc$^3$ and
$V_\mathrm{B}\approx1.8\times 10^4$ kpc$^3$, and the density profiles described
in section 2.2, we find $\overline{n_\mathrm{A}}\approx 1.7\times 10^5$
subdwarfs kpc$^{-3}$ and $\overline{n_\mathrm{B}}\approx 1.2\times 10^3$
subdwarfs kpc$^{-3}$.

Also plotted in Fig.~\ref{schematic} is the line of sight (for simplicity
depicted as a cylinder) for which surface photometry indicates anomalously red
colours in the stacked halo data of Z04. This region is located at a projected
distance of $R_\mathrm{proj}=2\times r_\mathrm{exp}$ from the midplane of the
disk, which -- when rescaled to a galaxy with the dimensions adopted for the
Milky Way -- corresponds to $R_\mathrm{proj}=5$ kpc. This sightline may or may
not directly intersect the volume B probed by deep-field star counts,
depending on the (undetermined) orientation of the Sun with respect to the surface
brightness sightline. However, in the axisymmetric halo models considered here,
this is of no consequence. The important point is instead that, because of the
projected nature of the surface photometry data, contributions from the light
measured along the depicted sightline can come from regions outside (i.e. in
front of, or behind) the \citet{Gould et al.} cone. Therefore, the available
star counts do not necessarily dictate what would be observed along the surface
photometry sightline, although they do set a lower limit on the total surface
brightness.

\subsection{Rejection criteria}
When testing red halo models against the observational constraints, we assume
that the red halo is smooth and can be described by a density profile of the
form:

\begin{equation}
n_\mathrm{red}(x,y,z)=n_\mathrm{red,\odot}\left(
\frac{x^2+y^2+(z/q_\mathrm{red})^2}{R_0^2} \right)^{-\beta_\mathrm{red}/2}
\label{n_red}
\end{equation}

where $x$, $y$, $z$ are Galactocentric coordinates. $n_\mathrm{red}$ is the number density of subdwarfs, $q_\mathrm{red}$ the
flattening parameter for the red halo and $\beta_\mathrm{red}$ the exponent of
the red halo density power-law. This matches the assumptions used in the work
of \citet{Gould et al.}.

For each combination of red halo model parameters considered, we calculate the
expected mean number densities of red halo subdwarfs in volumes A and B:

\begin{equation}
\overline{n_\mathrm{red,\: A/B}}=\frac{ }{V_\mathrm{A/B}}\int
n_\mathrm{red}(x,y,z) \; \mathrm{d}V_\mathrm{A/B}.
\end{equation}

All red halo models that give either $\overline{n_\mathrm{red,\:
A}}>\overline{n_\mathrm{A}}$ or $\overline{n_\mathrm{red,\:
B}}>\overline{n_\mathrm{B}}$ are then considered rejected. Given the red halo
density profile parameters $q_\mathrm{red}$ and $\beta_\mathrm{red}$, the
normalization parameter $n_\mathrm{red,\odot}$ is computed from the requirement
that surface brightness of the red halo, at a projected distance from the Milky
Way disk of $2\times r_\mathrm{exp}$ is equal to the observed
$I_{\mathrm{red},\; i}$, i.e. $\mu_{\mathrm{red},\; i}$ converted from mag
arcsec$^{-2}$ to $L_\mathrm{AB}$ kpc$^{-2}$:

\begin{equation}
I_{\mathrm{red},\; i} = \frac{2 n_\mathrm{red,\odot}}{N_\mathrm{sub}/L_i}
\int_0^{y_\mathrm{max}} \left(
\frac{y^2+(\frac{2r_\mathrm{exp}}{q_\mathrm{red}})^2}{R_0^2}
\right)^{-\frac{\beta_\mathrm{red}}{2}} \; \mathrm{d}y.
\label{Ieq}
\end{equation}

Here $N_\mathrm{sub}/L_i$ describes the number of subdwarfs per luminosity (in
units of $L_{\mathrm{AB},i}$, i.e. the $i$-band luminosity of the flat-spectrum
source used to define the zero point of the SDSS AB system) in the red halo
population. The upper integration limit is given by
$y_\mathrm{max}=\sqrt{r_\mathrm{max}^2-4r_\mathrm{exp}^2}$, where
$r_\mathrm{max}$ is the outer truncation radius of the red halo. To allow any
putative red halo maximum leverage, we set $r_\mathrm{max}$ equal to the virial
radius of the dark matter halo, which we here take to be 258 kpc \citep{Klypin
et al.}.

This leads to four free parameters for the red halo model: $q_\mathrm{red}$,
$\beta_\mathrm{red}$, $N_\mathrm{sub}/L_i$ and $\mu_{\mathrm{red},\; i}$.

For red halo models that remain viable after constraints A and B
have been imposed, we compute the mass contained in such structures using:

\begin{equation}
M_\mathrm{red}=\left(\frac{M}{N_\mathrm{sub}}\right)\int_{0}^{2\pi} \!\!\!
\int_{0}^{\pi} \!\!\! \int_{r_\mathrm{min}}^{r_\mathrm{max}} n(r,\theta,\phi)
r^2\sin(\theta) \; \mathrm{d}r \mathrm{d}\theta \mathrm{d}\phi,
\label{Meq}
\end{equation}

where $M/N_\mathrm{sub}$ represents the ratio of total mass of the red halo
stellar population (in the 0.08--120 $M_\odot$ mass range) to the number of
such stars considered subdwarfs, and $n(r,\theta,\phi)$ is simply equation
(\ref{n_red}) converted into spherical coordinates. Since we have assumed a
core-free, power-law density for the red halo (eq. \ref{n_red}), a non-zero
lower radial integration limit $r_\mathrm{min}$ is required to prevent
$M_\mathrm{red}$ from diverging for $\beta_\mathrm{red}>0$. Here, we have
adopted $r_\mathrm{min}=1$ kpc. Allowing a smaller $r_\mathrm{min}$ would only have
a significant impact on $M_\mathrm{red}$ for very steep density profiles
(i.e. profiles with high $\beta_\mathrm{red}$), since these attain very high
central densities. However, such populations are of little interest for the
missing-baryon problem, since models that attempt to hide a substantial
baryonic mass in the form of stars within $\sim 1$ kpc from the Milky Way
centre are subject to very strong constraints by microlensing observations
towards the bulge \citep{Calchi Novati et al. b}.
\begin{deluxetable}{lll}
\tabletypesize{\scriptsize}
\tablecaption{$N_\mathrm{sub}/L_i$ and $M/N_\mathrm{sub}$ as a function of IMF
slope $\alpha$.}
\tablewidth{0pt}
\tablehead{
\colhead{$\alpha$} & \colhead{$N_\mathrm{sub}/L_i$} & $M/N_\mathrm{sub}$ }
\startdata
0.75/2.35\tablenotemark{a} & 67 & 1.7\\
2.35 & 300 & 0.59 \\
3.00 & 650 & 0.60 \\
3.50 & 1100 & 0.70 \\
4.00 & 1800 & 0.86\\
4.50 & 2400 & 1.1\\
5.00 & 2700 & 1.4\\
\enddata
\tablenotetext{a}{This entry represents the two-component power-law IMF
considered representative for the standard stellar halo of the Milky Way.}
\label{NL_table}
\end{deluxetable}

\subsection{Spectral synthesis}
To determine reasonable values for the parameter $N_\mathrm{sub}/L_i$ in
equation (\ref{Ieq}), spectral synthesis modelling is required. The
constraints presented in Section 4 are based on the assumption that the red
halo has an age of 10 Gyr, a metallicity of $Z=0.001$, and a power-law IMF with
exponent $\alpha$ ($\mathrm{d}N/\mathrm{d}M \propto M^{-\alpha}$) throughout
the mass range 0.08--120 $M_\odot$. The star formation rate (SFR) is assumed to
have been exponentially declining over cosmological time scales
($\mathrm{SFR}\propto \exp(-t/\tau)$), with $\tau=1$ Gyr (suitable for an
early-type system). Since the resulting constraints are somewhat sensitive to
the parameter values adopted, the effects of relaxing these assumptions are
carefully explored in Section 5.

To derive $N_\mathrm{sub}/L_i$, we also need to identify the subset of stars in
the red halo population that qualify as subdwarfs. The luminosity criteria
described in section 2.2 convert into a stellar mass range of $0.15 \lesssim M
(M_\odot)\lesssim 0.7$ \citep[e.g.][]{Marigo et al.}.  The number of subdwarfs
$N_\mathrm{sub}$ within a population can then straightforwardly be derived from
the IMF, whereas the integrated $i$-band luminosity and the stellar population
mass $M$ can be derived using a spectral synthesis code. Here, we use the
population synthesis model PEGASE.2 \citep{Fioc & Rocca-Volmerange} to derive
these quantities.

The constraints on red halo models derived in section 4 will be computed for the
$N_\mathrm{sub}/L_i$ values (in units of dwarfs $L_{\mathrm{AB},i}^{-1}$)
listed in Table.~\ref{NL_table}. These correspond to the predictions for IMF
slopes $\alpha = 2.35$ (i.e. the Salpeter IMF), 3.00, 3.50, 4.00, 4.50
\citep[the value favoured by][]{Zackrisson et al. a} and 5.00. For comparison,
we also list $N_\mathrm{sub}/L_i$ for an IMF that more closely resembles that
of the hitherto detected stellar halo: a broken power-law IMF with
$\alpha=0.75$ in the 0.08--0.7 $M_\odot$ mass range and $\alpha=2.35$ for 0.7--120
$M_\odot$. This choice (labeled 0.75/2.35 in Table 1) is in fair agreement with
the mass function derived by \citet{Gould et al.}, and produces results similar
to other parameterizations of the halo IMF \citep[e.g.][]{Chabrier}. The
$M/N_\mathrm{sub}$ ratios, required to compute the stellar population masses of
viable red halo models using equation (\ref{Meq}), are also listed for these
IMFs.

These IMFs cover the range from perfectly normal to extremely bottom-heavy, and therefore account for the possibility that the $r-i$ colour (the primary reason for advocating a steep IMF slope) reported by Z04 may have been artificially reddened by PSF effects \citep{de Jong}.  

\begin{figure}[t]
\plotone{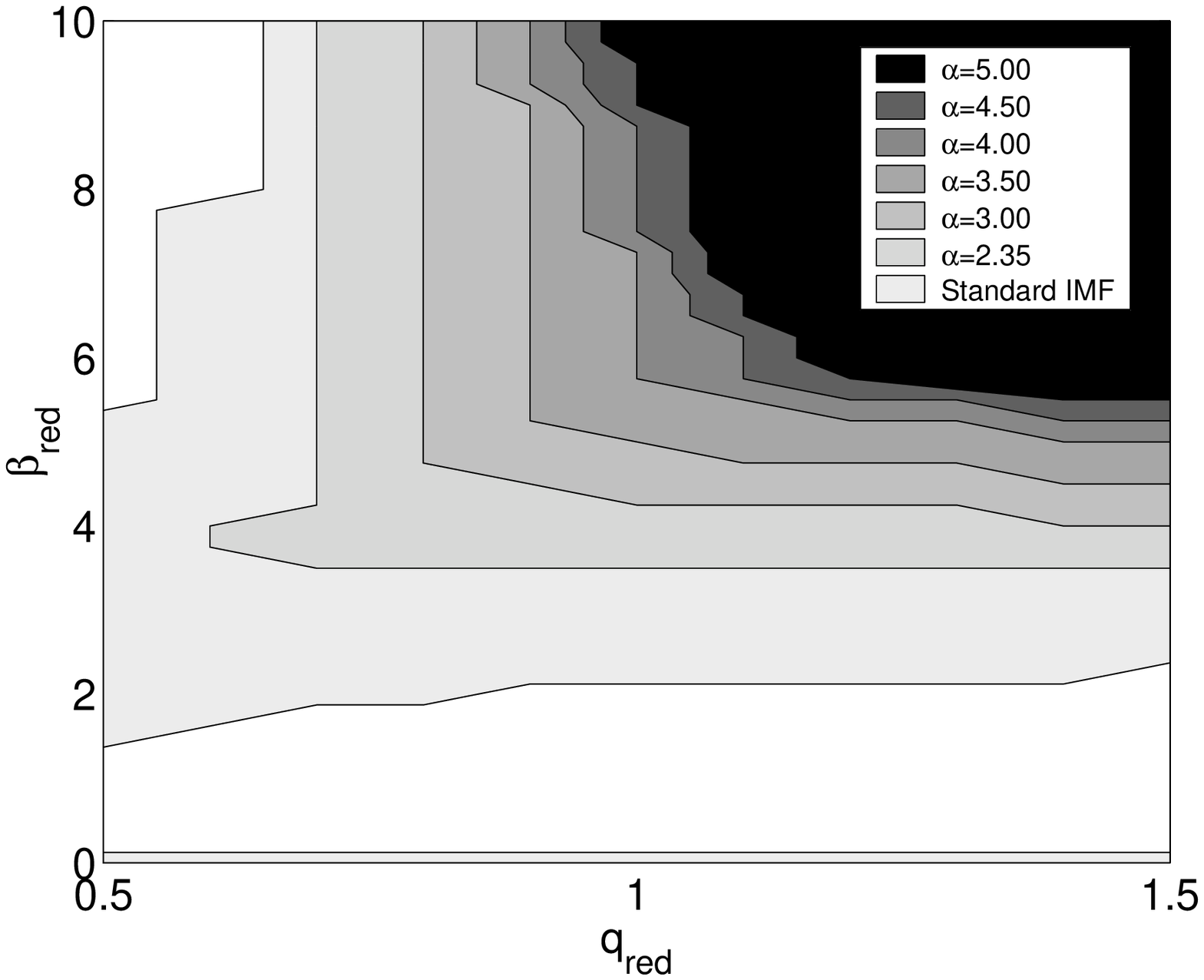}
\caption{The regions in the ($\beta_\mathrm{red}$,$q_\mathrm{red}$) parameter
space allowed for different red halo IMFs (shades of gray) in the case of
$\mu_{\mathrm{red},\; i}=26.7$ mag arcsec$^{-2}$. The constraints are based on the \citet{Digby et al.} and \citet{Gould et al.} surveys combined -- i.e. red halos are considered rejected if either $\overline{n_\mathrm{red,\:
A}}>\overline{n_\mathrm{A}}$ (conflict with \citealt{Digby et al.}) or $\overline{n_\mathrm{red,\:B}}>\overline{n_\mathrm{B}}$ (conflict with \citealt{Gould et al.}). The ``Standard IMF'' refers to the broken power-law IMF considered representative of the hitherto detected stellar halo of the Milky Way. Darker areas (corresponding to more bottom-heavy IMFs) are here plotted on top of brighter ones, so that the region marked by the darkest shade of gray indicates the boundaries inside which halos of {\it any} IMF (ranging from the standard IMF to the $\alpha=5.00$ IMF) are allowed to lie. Halo models with the standard IMF are on the other hand only constrained to lie within the union of all the differently shaded areas. The ($q_\mathrm{red}$, $\beta_\mathrm{red}$) constraints on halo models with less extreme IMFs (lighter shades of gray) are consequently much weaker than those of the most bottom-heavy ones (darker shades of gray). White regions represent regions of the parameter space where no red halo models are allowed.}
\label{paramspace}
\end{figure}
\begin{figure*}[t]
\plottwo{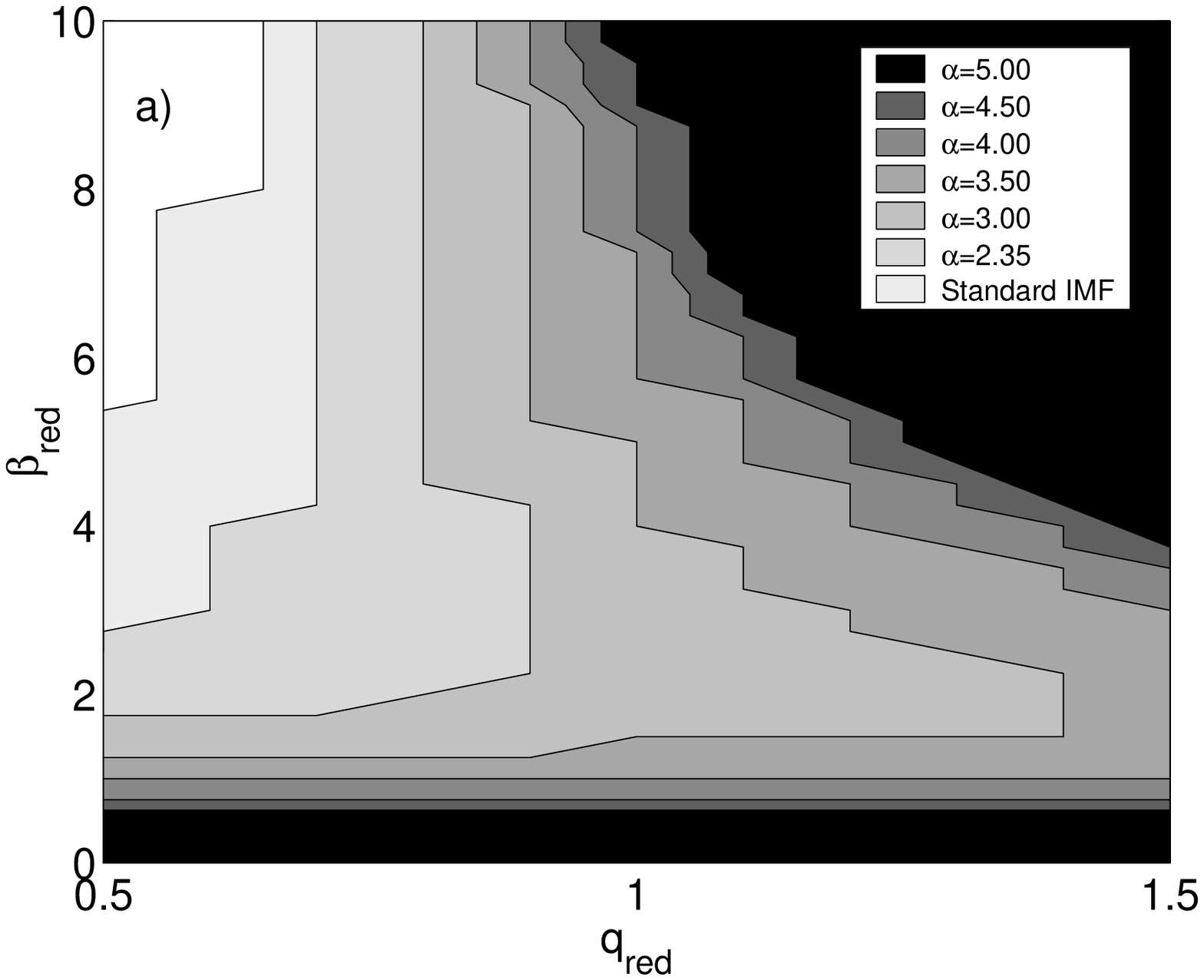}{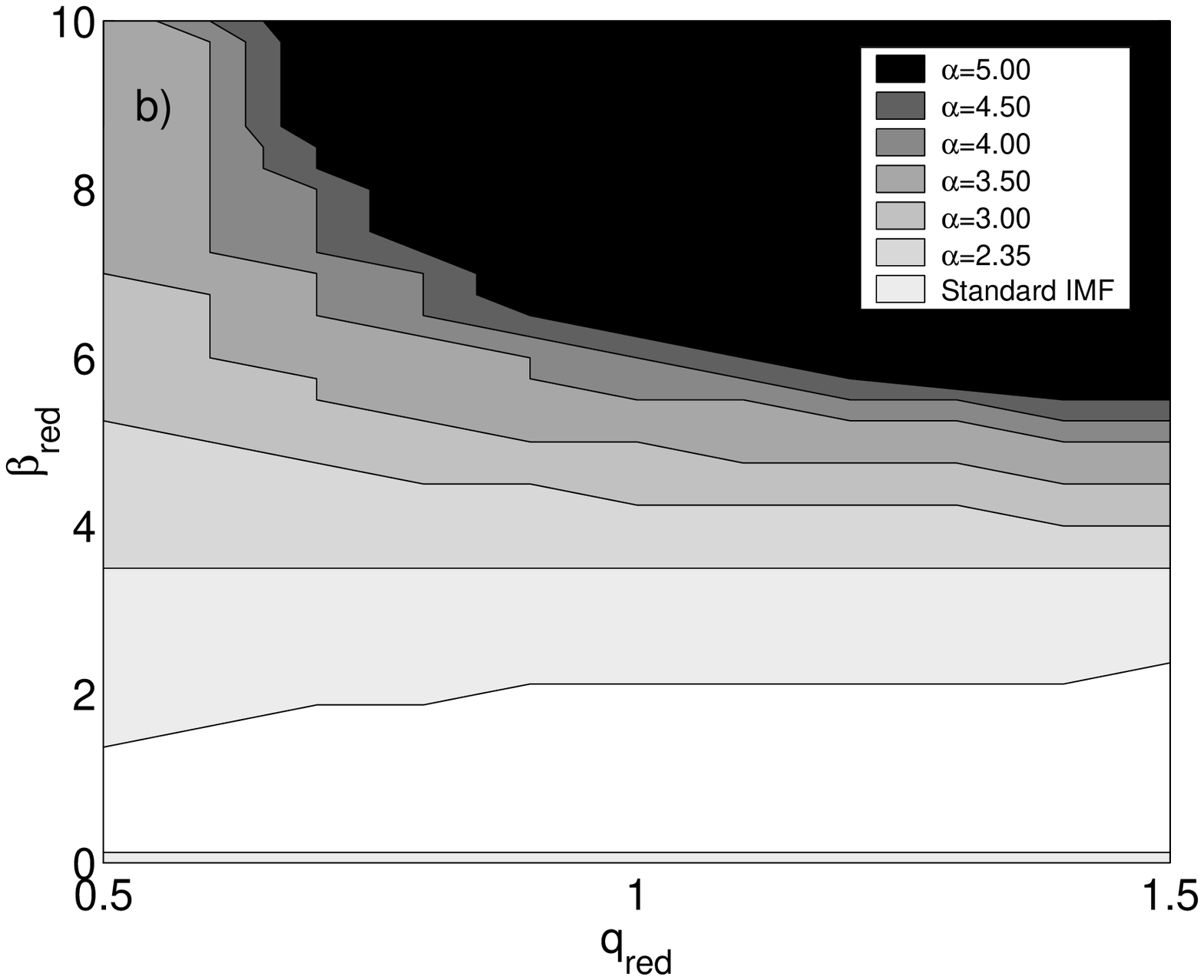}
\caption{Same as Fig.~\ref{paramspace}, but showing the regions of the ($\beta_\mathrm{red}$,$q_\mathrm{red}$) red halo parameter space allowed by the data sets of ({\bf a}) \citet{Digby et al.} and ({\bf b}) \citet{Gould et al.}, separately.}
\label{paramspace2}
\end{figure*}
 
\section{Constraints on red halo models}
By simply adopting the best-fitting structural parameters derived by Z04
($q_\mathrm{red}\approx 0.6$, $\beta_\mathrm{red}\approx 3$) and assuming a
bottom-heavy IMF population with $\alpha=4.50$ for the red halo colours, we
would arrive at a red halo mass of $M_\mathrm{red}\approx 2\times 10^{10}\
M_\odot$. While this is in the right ballpark for accounting for a significant
fraction of the missing baryons in the Milky Way, this parameter combination --
along with all similar ones -- are completely ruled out by the star
counts. This is demonstrated in Fig.~\ref{paramspace}, where we apply the combined constraints imposed by the \citet{Digby et al.} and \citet{Gould et al.} surveys to red halo models with parameters in the
range $\beta_\mathrm{red}=0$--10 (where $\beta_\mathrm{red}=0$ corresponds to a
constant-density halo) and $q_\mathrm{red}=0.5$--1.5. For these constraints, we
have assumed a scaling given by a halo surface brightness level of
$\mu_{\mathrm{red},\; i}=26.7$ mag arcsec$^{-2}$ (in direct correspondence to
the stacked halo by Z04). 

A wide range of different model parameters are allowed in the case of the
standard halo IMF (lightest shade of gray), including the $\beta=3$--3.5,
$q=0.5$--1.0 range in which the hitherto detected stellar halo is known to
lie. Of course, this IMF cannot account for the anomalously red colours of the
Z04 halo, and the mass of such structure is only about $1\times 10^9 \ M_\odot$, in fair agreement with estimated total mass of the ordinary stellar halo of the Milky Way \citep[e.g.][]{Bell et al.}. 

As the IMF becomes more bottom-heavy, the allowed region of the
parameter space is progressively pushed into the upper right corner of this
diagram. The most bottom-heavy IMFs considered (darkest shades of gray) are
only allowed at $q_\mathrm{red}>1$ (i.e. halos elongated in the polar
direction) and $\beta_\mathrm{red}>5$ (i.e. halos where the density drops much
faster as a function of distance from the centre than the standard halo with
$\beta\approx 3$). Because of their high central densities, such models more
closely resemble elongated bulges than normal halos.

The range of density profile slopes ($\beta_\mathrm{red}=2.5$--3.5) and flattenings
($q_\mathrm{red}=0.5$--0.7) favoured by Z04 are {\it completely ruled out} for
the Milky Way in the case of the more bottom-heavy IMF slopes. While there are
indeed red halo models ($\beta_\mathrm{red}>5$) with very bottom-heavy IMFs that would be able to
explain the $i$-band surface brightness while evading the constraints set by
current star counts, {\it the mass contained in such structures is very
low}. In the case of $\alpha=4.0$--5.0 \citep[i.e. an IMF similar to that advocated by][]{Zackrisson et al. a}, the maximum
red halo mass among the acceptable models is only $\approx 5\times 10^8\
M_\odot$, insufficient to be of any relevance for the missing-baryon problem.

Due to the different volumes probed by \citet{Digby et al.} and \citet{Gould et al.}, these two data sets constrain slightly different regions of the ($\beta_\mathrm{red}$,$q_\mathrm{red}$) parameter space. This is demonstrated in Fig.~\ref{paramspace2}, where the constraints imposed by \citet{Digby et al.} and \citet{Gould et al.} are plotted separately. While both sets of star counts are equally effective in ruling out the red halo models directly favoured by Z04 (i.e. $\beta_\mathrm{red}=2.5$--3.5 and $q_\mathrm{red}=0.5$--0.7), there are notable differences in the constraints imposed on more extreme models. Due to the smaller volume covered by \citet{Digby et al.}, their data are unable to rule out red halo models with close-to-constant densities (i.e. $\beta_\mathrm{red}\approx 0$). While such models imply relatively few subdwarfs in the vicinity of the Sun, the fact that these densities are retained far into the halo -- where the relevant volumes elements become huge -- result in very large red halo masses ($\sim 10^{12}\ M_\odot$). Since the \citet{Gould et al.} star counts are sensitive to subdwarfs at much greater distances from the Galactic centre, all $\beta_\mathrm{red}\approx 0$ models are rejected in Fig.~\ref{paramspace2}b. On the other hand, the position of the Digby et al. volume in the plane of the Milky Way imply that the these star counts are more sensitive than the the Gould et al. ones to red halo models which are flattened towards the disk (i.e. $q_\mathrm{red}<1.0$), and this is also the main contribution of Digby et al. to the combined constraints presented in Fig.~\ref{paramspace}.
\begin{figure}[t]
\plotone{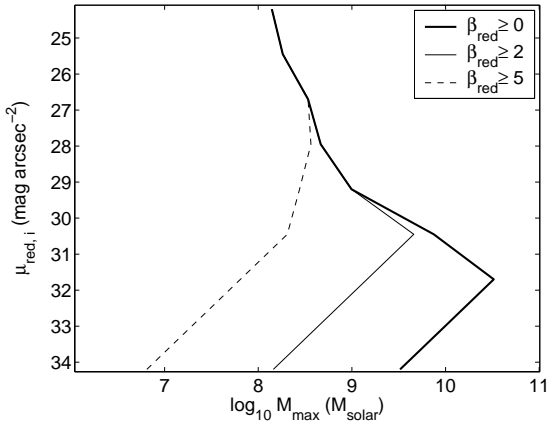}
\caption{Maximum allowed red halo mass vs. the adopted surface
brightness level $\mu_{\mathrm{red},\; i}$ at $r_\mathrm{proj}$ for a halo
population with IMF slope $\alpha=4.50$ and different constraints on the slope
of the red halo density profile: $\beta_\mathrm{red}\geq 0$ (thick solid line),
$\beta_\mathrm{red}\geq 2$ (thin solid), and $\beta_\mathrm{red}\geq 5$ (thin
dashed).}
\label{Mmax}
\end{figure}

\subsection{Varying the brightness of the halo}
In Fig.~\ref{Mmax}, we explore the maximum red halo mass allowed as a function
of the adopted $\mu_{\mathrm{red},\; i}$ in the case of the $\alpha=4.50$
IMF. Allowing any putative red halo around the Milky Way an $i$-band surface
brightness level brighter than $\mu_{\mathrm{red},\; i}=26.7$ mag arcsec$^{-2}$
has the effect of pushing the allowed parameter space for the bottom-heavy
models farther into the upper right corner of Fig.~\ref{paramspace}, i.e. to
higher $\beta_\mathrm{red}$ and $q_\mathrm{red}$. As a result, the largest red
halo masses among models with IMF slopes of $\alpha=4.0$--5.0 also become
somewhat smaller than those in the case of $\mu_{\mathrm{red},\; i}=26.7$ mag
arcsec$^{-2}$. This means that, if the red halo component is brighter in the
Milky Way than in the stacked halo, the red excess can only be attributed to a
stellar population with a bottom-heavy IMF if its density profile is much
steeper than that of both the Galactic stellar halo and the stacked halo of
external disk galaxies. Even if this were the case, the mass contained in this
structure would be orders of magnitude below that required to have any
relevance for the missing-baryon problem in the Milky Way.

As previously discussed, the $i$-band surface brightness of the halo may be contaminated by both a contribution from the disk and by the far wings of the PSF. Allowing a surface brightness level fainter than
$\mu_i=26.7$ mag arcsec$^{-2}$ increases the allowed parameter space for
bottom-heavy IMFs. At $\mu_{\mathrm{red},\; i}=29.2$ mag arcsec$^{-2}$ (i.e. a
factor of 10 fainter than the stacked halo), red halos with $\alpha=4.50$ and
structural parameters similar to the standard halo are allowed
(e.g. $\beta_\mathrm{red}=3.5$, $q_\mathrm{red}=0.75$). However, since this has
been made been possible at the expense of a low overall density scaling, the
maximum mass in such red halo models do not exceed $\sim 10^9\ M_\odot$, which is insufficient to explain the missing baryons.

With a red halo as faint as $\mu_{\mathrm{red},\; i}=31.7$ mag arcsec$^{-2}$
(i.e. 100 times fainter than the stacked halo), the entire parameter space
depicted in Fig.~\ref{paramspace} becomes viable for all IMFs considered. As a
result, red halos models with $\alpha=4$--5 and masses in the range that would
indeed be relevant for the missing-baryon problem (several times $10^{10}\
M_\odot$) evade the observational constraints considered here. However, as
shown in Fig.~\ref{Mmax} these high masses are produced exclusively by models
with $2>\beta_\mathrm{red}\geq 0$. Density profile slopes like these are too shallow to be consistent with
the slopes of $\beta_\mathrm{red}=2.5$--3.5 favoured by Z04. Since Z04 made no
attempt to correct their surface brightness profiles for PSF effects, it is
moreover likely that their estimate of $\beta_\mathrm{red}$ is an underestimate
rather than an overestimate.
 
As is evident from Fig.~\ref{Mmax}, the maximum red halo mass reaches a peak at
some $\mu_{\mathrm{red},i}$ and then declines for fainter models, since the
allowed region of the $(\beta_\mathrm{red}$, $q_\mathrm{red})$ plane initially
grows as fainter halos are considered. However, once the entire plane becomes
permitted, the only effect of going fainter is to lower the overall density
scaling of the red halos, thereby giving smaller total masses.

Hence, while a red halo of low-mass stars can in principle evade the current
constraints on halo subdwarfs and at the same time account for some
non-negligible fraction of the missing baryons, this would require that {\it both}
the scaling and slope of its surface brightness profile would be very different
from that of the halo seen around stacked external galaxies. Since the variance
of red halo properties among disk galaxies cannot easily be assessed from the
Z04 study, this possibility cannot be ruled out. However, advocating a solution
of that type would relieve the stacked halo of all predictive power concerning
the Milky Way, and we do not consider this possibility any further.

Allowing for the fact that the $r-i$ colour of the halo may have been overestimated (thereby implying a 
less extreme halo IMF) does not significantly alter these conclusions. All halo models with $\beta_\mathrm{red}\geq 2$ that evade the star counts constraints have $M_\mathrm{max}\leq 10^{10}\ M_\odot$ at all $\mu_i$ considered in Fig.~\ref{Mmax}, regardless of which of the IMFs in Table 1 we adopt. 

\section{Discussion}
Our results indicate that a smooth halo with a bottom-heavy IMF and structural parameters similar to those of the stacked halo is completely ruled out in the Milky Way by current star count data. A halo component with a bottom-heavy IMF
would have to have an overall scaling or density profile that differs
substantially from that of the stacked halo to remain viable. Moreover, all
permitted smooth red halo models with a density slope {\it even remotely similar} to
that of the stacked halo contain far too little mass to have any bearing on the
missing-baryon problem in the Milky Way.

These conclusions are admittedly based on a large number of assumptions
regarding the properties of the red halo. Since one can in principle consider
many formation scenarios for a halo-like structure dominated by low-mass stars
-- for example in situ formation, population III stars, dynamical mass segregation
\citep[see][for a more details discussion]{Zackrisson et al. b} -- the
properties of the red halo population may be very different from those thus far
adopted. Here, we therefore investigate the robustness of our conclusions in
light of the most important assumptions made.

\subsection{Age, star formation history and metallicity of the halo}
\begin{figure*}
\plottwo{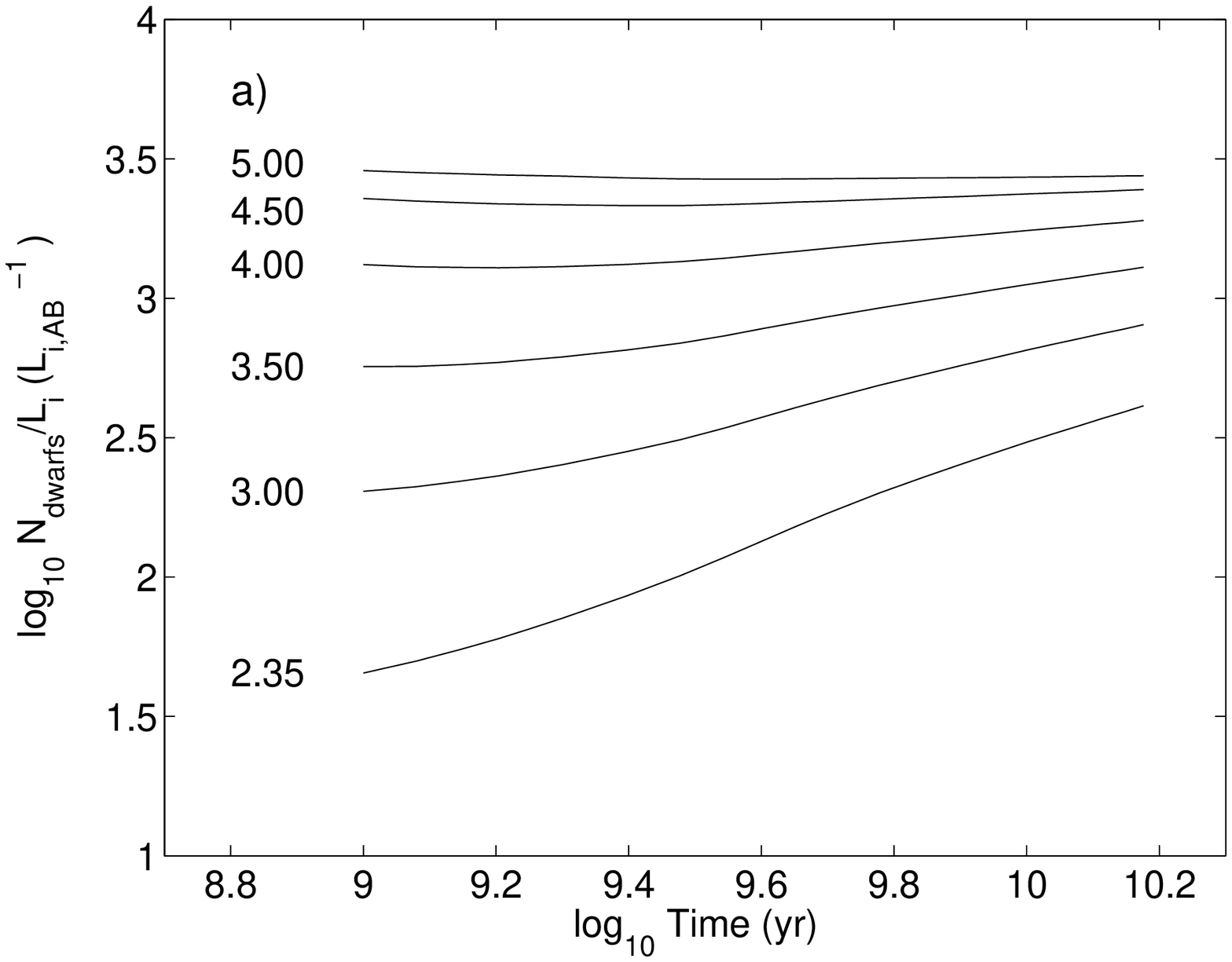}{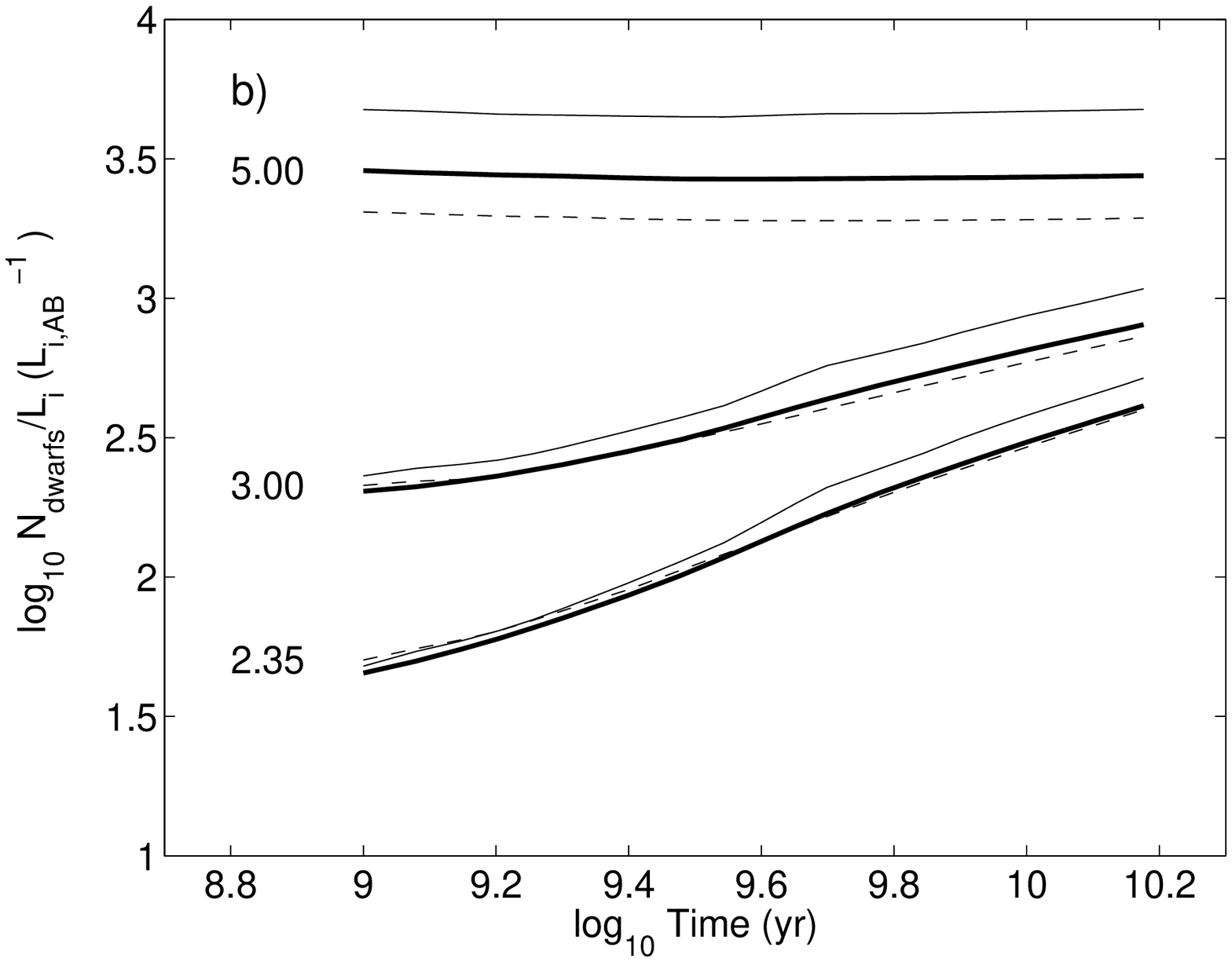}
\caption{$N_\mathrm{sub}/L_i$ as a function of age for stellar populations with
different IMF slopes $\alpha$ (indicated by labels).  {\bf a)} Populations with
$Z=0.001$ and $\tau=1$ Gyr. {\bf b)} Populations with $\tau=1$ Gyr at
metallicities of $Z=0.0001$ (thin dashed), $Z=0.001$ (thick solid) and $Z=0.008$
(thin solid). To avoid cluttering, only a subset of IMF slopes are
included. \label{N/L}}
\end{figure*}
In deriving the constraints presented in section 4, fixed $N_\mathrm{sub}/L_i$
ratios have been adopted for every IMF slope considered, based on the assumed
age, prior star formation history and metallicity of the red halo. In reality,
of course, we have but very poor constraints on these properties. For a halo
population, one naively expects a relatively high age and a low metallicity,
and we therefore restrict the discussion to ages $t\geq 1$ Gyr and
metallicities $Z\leq 0.008$. In Fig.~\ref{N/L}a and b, we demonstrate the
dependence of $N_\mathrm{sub}/L_i$ on age and $Z$ in the case of a stellar
population with $\tau=1$ Gyr. The $N_\mathrm{sub}/L_i$ ratio generally increases as a
function of age.

However, for the more bottom-heavy IMFs (i.e. high IMF slope $\alpha$), the age
dependence is weak, and an age lower than the 10 Gyr assumed in section 4 would
allow $N_\mathrm{sub}/L_i$ to be lowered by no more than $\approx 25\%$
(for $\alpha\geq 4$). Allowing a metallicity higher than the assumed value of
$Z=0.001$ would just increase $N_\mathrm{sub}/L_i$ and hence strengthen the
constraints. A metallicity as low as $Z=0.0001$ would decrease
$N_\mathrm{sub}/L_i$, but by no more than $\approx 30\%$. The maximum decrease is
moreover attained at the highest ages, which means that this drop in
$N_\mathrm{sub}/L_i$ does not add to the age effect -- you cannot have both at
the same time. Hence, the combined effect of age and metallicity is no more
than $\approx 30\%$ for IMFs with $\alpha\geq 4$.

Alterations in the star formation history affect the $N_\mathrm{sub}/L_i$ ratio
in a way that is very similar to age variations. The maximised
$N_\mathrm{sub}/L_i$ are attained in the case of an instantaneous burst of star
formation (i.e. a single-age population). While the perfectly coeval onset and
quenching of star formation associated with this scenario seems unrealistic
given the huge spatial scales involved in the halo, an instantaneous burst
would just make the constraints on the red halo stronger. Allowing a
more prolonged star formation history than the $\tau=1$ Gyr adopted in section
4 on the other hand lowers $N_\mathrm{sub}/L_i$. In the extreme case of having
a star formation that increases as a function of time, the
$N_\mathrm{sub}/L_i$ ratio stays almost constant at the value attained at the lowest
ages. Hence, such scenarios do not provide the means for lowering
$N_\mathrm{sub}/L_i$ more than the age effects already discussed.

In summary, $N_\mathrm{sub}/L_i$ for these alternative scenarios may be
somewhat lower ($\approx 30\%$) than the values listed in Table \ref{NL_table}
and used in the constraints presented in section 4. A detailed inspection of
the entries in Table \ref{NL_table} reveals that this may shift the constraint
levels by at most one level, in the sense that the constraints given for an IMF
with slope $\alpha$ may relaxed to mimic those for an IMF with a slope one step
lower among those tested. As an example, changing the age from 10 to 1 Gyr
for an $\alpha=4.00$ model would lower $N_\mathrm{sub}/L_i$ from
$N_\mathrm{sub}/L_i\approx 1800$ to $\approx 1300$ (see Fig.~\ref{N/L}a),
i.e. a decrease by $\approx 30\%$. This $N_\mathrm{sub}/L_i$ ratio is
intermediate between those listed for $\alpha=3.50$ and 4.00. Hence,
the relaxed constraints on the allowed structural parameters for a
$\alpha=4.00$ halo would fall between those of the $\alpha=3.50$ and
4.00 contours in Fig.~\ref{paramspace}. This shift is too small to
allow a halo population with $\alpha=4.0$--5.0 into regions of the
($q_\mathrm{red}$, $\beta_\mathrm{red}$) parameter space where it would
contribute significantly to the missing-baryon problem.  Other spectral
synthesis codes may produce slightly different values for $N_\mathrm{sub}/L_i$,
but the qualitative impact of such variations on the final constraints is easy
to assess from Table~\ref{NL_table} and Fig.~\ref{N/L}.

Once the constraints on $q_\mathrm{red}$ and $\beta_\mathrm{red}$ have been
determined, the mass of the viable red halo models are computed using equation (\ref{Meq}), based on the $M/N_\mathrm{sub}$ ratio. This ratio also has a
slight dependence on age and star formation history, but varies by no more than
$\approx 30\%$ for IMFs with slopes $\alpha\geq 3$. The inferred red halo
masses change insignificantly because of such effects.

Hence, we argue that our conclusions are robust with respect to the assumptions
made about the age, metallicity and star formation history of the red halo.

\subsection{Disk scale length}
The constraints presented in section 4 have been derived using a scale length
for the Milky Way of $r_\mathrm{exp}=2.5$ kpc. The effect of changing
$r_\mathrm{exp}$ is to spatially shift the projected distance
$r_\mathrm{proj}=2\times r_\mathrm{exp}$ between the plane of the disk and the
region where the density scale of the red halo models are set by the
assumed $\mu_{\mathrm{red},\; i}$. Adopting a smaller value of $r_\mathrm{exp}$ weakens the
constraints imposed on red halo models by star counts in volumes A and B, whereas a larger $r_\mathrm{exp}$ would make the
constraints stronger. Allowing $r_\mathrm{exp}=2.0$ kpc \citep[at the lower
limit of what is allowed by the constraints set by][]{Juric et al.} only allows
an increase of the overall mass of red halos with IMFs with
$\alpha=4.0$--5.0 by a factor of $\approx 2$, which still places the red halo
mass substantially below the range relevant for the missing-baryon problem in
the Milky Way. Therefore, our conclusions seem very robust with respect to
uncertainties in the scale length of the Milky Way disk.

\subsection{Halo IMF}
So far, we have only considered power-law IMFs with single-valued slopes
$\alpha$ for the red halo population. In principle, far more complicated IMFs
(e.g. broken power laws, lognormal IMFs, IMFs with mass spikes) could be
considered. While this is not warranted given the large observational
uncertainties associated with current red halo data, improved measurements may
indeed require modifications of the assumed IMF. The revised constraints can be
assessed by comparing $N_\mathrm{sub}/L_i$ and $M/N_\mathrm{sub}$ of
populations with more complicated IMFs to the values listed in
Table~\ref{NL_table}. However, as long as the IMF is not tweaked to move a very large fraction of the overall population mass outside the stellar mass range to which the current subdwarf star counts are sensitive ($\approx 0.15-0.7 \ M_\odot$), the
conclusions presented here will not be qualitatively altered.

\subsection{Halo density profile}

The constraints presented are based on the reasonable assumption that the
volume density of red halo stars decreases monotonically with distance from the
Galactic centre. Shells and tidal tails are sometimes seen in the outskirts of
galaxies, and represent overdensities of matter that formally violate this
assumption. If one entertained the notion that the abnormal colours of the red
halo come from low-mass stars in a shell-like structure, then the constraints
presented here can in principle be completely sidestepped by pushing the radius
of this shell to distances outside the volume probed by the \citet{Gould et
al.} star counts (volume B in Fig.~\ref{schematic}). However, while the
density of such a structure can be chosen freely to correspond to the measured
surface brightness level at any single projected radius, the decrease in
surface brightness as a function of projected distance from the centre would be
far slower than that reported by Z04.

So far, we have assumed the red halo density profile to be a single-valued
power law, whereas more complicated density profiles can of course be
considered. A broken power-law profile with a shallow slope (or a constant
density core) close to the centre ($R_\mathrm{c}< R_\mathrm{proj}$) would
result in constraints on the outer slope and flattening identical to those
presented here, except that the overall red halo mass would become even lower
than implied by current constraints for the allowed parameter combinations. The
situation becomes more complicated if the power-law density profile changes
slope inside (or even outside) the volumes A and B. However, a break of this
type would be imprinted in the red halo surface brightness profiles at some
level, and there is no compelling evidence for such features at the current
time.

\subsection{Subdwarf detectability as a function of distance}

The luminosity functions of \citet{Digby et al.} and \citet{Gould et al.} have
for simplicity been adopted throughout volumes A and B for all subdwarfs in
the luminosity range $7.5\lesssim M_V\lesssim 12.4$. In reality, the luminosity
function is not sampled equally well at all distances, since the faintest
(i.e. least massive) subdwarfs typically cannot be detected out to as large
distances as the brightest ones. In fact, $M_V\approx 12.4$ stars are detected
only out to a distances of order 5 kpc from the Sun in volume B,
i.e. substantially smaller than the 40 kpc adopted for its outer boundary. The
constraints imposed by these volumes may therefore be too strict for red halo
populations composed almost entirely of the subdwarfs at the lowest
masses. However, even if we completely disregard all constraints based on
volume B, thereby restricting ourselves to very small distances from the Sun
($\leq 2.8$ kpc), we are still left with reasonably strong constraints imposed by volume A (as depicted in Fig.~\ref{paramspace2}a). All
$\alpha=4.50$ populations with $\beta_\mathrm{red}\geq 2.75$ are constrained to
contribute less than $10^{10}\ M_\odot$ to the baryon budget for all $\mu_i$
normalizations considered. Shallower density profiles $\beta_\mathrm{red}<
2.75$ do allow red halo masses in excess of $10^{10}\ M_\odot$, but these would
be in poor agreement with the values derived by Z04.

\subsection{Halo smoothness}

One way of sidestepping the Milky Way star counts would be to assume that
the red halo stars are not smoothly distributed, but clustered. Scenarios of
this type have been carefully investigated by \citet{Kerins a,Kerins b}, and
remain a viable way of hiding the red halo population from detection in the
Milky Way. The volumes A and B in which the star counts have
assumed to be valid (Fig.~\ref{schematic}), are crude representations of the
actual volumes probed. In reality, the subdwarf samples used in the studies by
\citet{Gould et al.} and \citet{Digby et al.} have been obtained in a limited
number of fields, which would more accurately be represented by a large number
of narrow cones probing the volume. If red halo stars were clustered, we may
have missed them simply because they have so far happened to fall between the
cones.

Such clusters of low-mass stars in the Milky Way halo might already be
detectable as faint, extended objects in a wide-angle survey such as the SDSS
(provided that a sufficient number of such clusters is located at sufficiently
small distances from us), or in external galaxies, using a combination of
surface photometry and star counts.

As an example, consider a star cluster of mass $4\times 10^4 \ M_\odot$
\citep[the mass favoured by][]{Kerins a,Kerins b} with $Z=0.004$ and IMF slope
$\alpha=4.50$ \citep[i.e. the values favoured by][]{Zackrisson et al. a}. The
cluster would have an absolute magnitude of $M_I\approx -3.3$ in the Cousins
$I$-band, which is $\approx 0.5$ mag below the tip of the red giant branch
(RGB) in a 10 Gyr old population. Objects this bright are readily detectable
with the HST at distances out to at least $\approx 10$ Mpc. However, objects of
this type could possibly -- depending on their compactness and exact distance --
be mistaken for single stars when observed in the commonly used $V$ and $I$
filters alone, since the $V-I$ colour of such a population is predicted to be
$V-I\approx 1.7$ (assuming $\tau=1$ Gyr and an age of 10 Gyr), similar to that
of RGB stars. Multicolour data should nonetheless be able to reveal their
exotic nature, since the spectra of integrated stellar populations typically differ
substantially from those of individual stars.

The idea of red halo star clusters at these luminosities is very interesting in
the light of recent work by \citet{Yan et al.}, who in the halo of the nearby
galaxy M60 (distance $\approx 16$ Mpc) may have detected a population of halo
objects with luminosities of giants, but $izJHK$ colours that are
difficult to reconcile with current models of such stars. Yan et al. argue that
these objects, whatever their nature, may be responsible for the anomalous
colours of red halos. It remains to be investigated, however, whether the
colours of these objects may be consistent with the expectations for
star clusters with bottom-heavy IMFs.

\subsection{The typicality of the Milky Way}

Since the analysis by Z04 does not directly reveal what the variance of halo
properties within the stacked sample is, we do not know what fraction of the
stacked galaxies have a red halo. Perhaps some do not, and the Milky Way is
just one such case. If so, we are confronted with red halos possibly
contributing significantly to the baryonic mass in some disk galaxies -- but
not in others. In disks at least, the baryonic Tully-Fisher relation
(i.e. the relation between the total inferred stellar and gas masses vs. the
rotational velocity) can be tuned via the stellar population mass-to-light
ratio to show very small scatter \citep[][]{McGaugh et al.,McGaugh}, which has
been used as an argument that most of the baryons of disk galaxies have already been
identified. Having considerable baryonic reservoirs in red halos
for some disk galaxies, but not for others, would supposedly increase the scatter, but the magnitude of this effect is unclear. Moreover, there are observational indications that the Milky Way may be offset from the baryonic Tully-Fisher relation \citep{Flynn et al.}. If this is the case, constraints based on the Milky Way may not be able to say anything
definitive on the red halo masses of disk galaxies in general.

\section{Summary}
By comparing the photometric properties of the red halo detected in stacked
SDSS data by Z04 to the available subdwarf star counts in the Milky Way halo,
we have tested the viability of models that aim to explain the colours of red
halos as due to stellar populations with abnormally high fractions of low-mass
stars. We find, that a smooth halo with a bottom-heavy IMF and structural
parameters similar to those of the stacked halo is completely ruled out in the
Milky Way by current star count data. A halo component with a bottom-heavy IMF
would have to have an overall scaling or density profile that differs
substantially from that of the stacked halo to remain viable. Moreover, all
permitted smooth red halo models with a density slope even remotely similar to
that of the stacked halo contain far too little mass to have any bearing on the
missing-baryon problem in the Milky Way. These conclusions could possibly be
avoided if the red halo stars are locked up in small star clusters. We argue
that such scenarios can be tested through combined star counts and deep surface
photometry of the halos of nearby galaxies.

\acknowledgments
EZ acknowledges research grants from the Academy of
Finland, the Swedish Research Council and the Swedish Royal Academy of
Sciences. The anonymous referee is thanked for useful comments which helped improve the quality of the paper.

\end{document}